\documentclass[a4paper,11pt]{JHEP3}
\usepackage[active]{srcltx}
\JHEPspecialurl{http://jhep.sissa.it/JOURNAL/JHEP3.tar.gz}
\usepackage{epsfig,multicol,bbm,amsmath,times,graphicx,amssymb}
\usepackage{dcolumn}
\usepackage{bm}
\usepackage{cite}

\newcommand{\bea}{\begin{eqnarray}}
\newcommand{\eea}{\end{eqnarray}}
\newcommand{\bean}{\begin{eqnarray*}}
\newcommand{\eean}{\end{eqnarray*}}
\newcommand{\nn}{\nonumber \\}

\def\W #1{\widetilde{#1}}

\def\a{{\alpha}}

\def\b{{\beta}}

\def\Label#1{\label{#1}%
  \smash{\hbox to0pt{\raise1ex\hbox{\tiny[#1]}\hss}}}

\title{Proof of Gravity and Yang-Mills Amplitude Relations}
\author{ N.E.J. Bjerrum-Bohr$^a$, Poul H. Damgaard$^a$, Bo Feng$^b$,
 {Thomas S{\o}ndergaard$^a$}\bigskip\\
{$^a$\small Niels Bohr International Academy and Discovery Center,
The Niels Bohr Institute, Blegdamsvej 17, DK-2100 Copenhagen,
Denmark\bigskip~\\$^b$\small Center of Mathematical Science,
Zhejiang University, Hangzhou, China \\ \small Kavli Institute for
Theoretical Physics China, CAS, Beijing 100190, China }}

\date{\today}

\received{\today} 		
\accepted{\today}		


\abstract{Using BCFW on-shell recursion techniques, we prove
a sequence of explicit $n$-point Kawai-Lewellen-Tye relations
between gravity and Yang-Mills amplitudes at tree level.}

\begin{document}

\section{Introduction}

One of the most challenging problems in theoretical physics is
the merging of quantum mechanics and gravity. Even at the most
modest level of perturbative quantum gravity at tree level this
is a daunting task. The Lagrangian of Einstein Gravity,
although compact in its formal expression, explodes into an
infinite series when expanded perturbatively around a given
background. Calculations of scattering amplitudes directly by
Feynman diagrams are virtually impossible, even at
tree level, for more than a few external gravitons. We must
therefore be more imaginative and try to look for alternatives.
The famous set of Kawai-Lewellen-Tye (KLT)
relations~\cite{Kawai:1985xq} provides just such an amazing
alternative. The KLT-relations express, in a highly surprising
manner, on-shell graviton amplitudes as `squares' of on-shell
color-ordered {\em Yang-Mills amplitudes}. If it were not for
the existence of string theory, such remarkable relations would
presumably never have been discovered. In string theory, where
they were originally derived~\cite{Kawai:1985xq}, they provide
what is by now a textbook example of the relationship between
open and closed string amplitudes. As such they are exact, and
when expanded they hold to all orders in $\alpha'$. Previously,
the only known way to deduce the corresponding amplitude
relations at the quantum field theory level was to take the
limit $\alpha' \to 0$ of these stringy
KLT-relations~\cite{Berends:1988zp,Elvang:2007sg}.

Inspired by Witten~\cite{WittenTopologicalString}, there has
been really remarkable progress in Yang-Mills amplitude
computations as reviewed in,
$e.g.$,~\cite{Cachazo:2005ga,Bern:2007dw}\footnote{ For a
review with a strong focus on KLT-relations
see~\cite{Bern:2002kj}.}. Very recently, we have shown how the
KLT-relations can be proven directly at the quantum field
theory level, without recourse to string
theory~\cite{BjerrumBohr:2010ta}. The method used in that paper
was the on-shell BCFW recursion technique~\cite{Britto:2004ap}.
It relies only on general quantum field theory properties of
the $S$-matrix and a certain convergence requirement on
amplitudes at complex momenta.

Much of the recent progress in the understanding of
KLT-relations is due to the discovery by Bern, Carrasco and
Johansson (BCJ) of a new set of gauge invariant relations
between color-ordered Yang-Mills amplitudes~\cite{Bern:2008qj}.
While these relations (and their generalizations to include
matter fields~\cite{Sondergaard:2009za}) were at first
conjectured on the basis of observed patterns in Yang-Mills
amplitudes, it is interesting that the first
proof~\cite{BjerrumBohr:2009rd} of their validity relied on
first shoving a more general set of relations in string theory
and then taking the $\alpha' \to 0$ limit, see also
ref.~\cite{Stieberger:2009hq}. In this way string theory again
plays a central role, and one can, using monodromy in string
theory, easily prove in this way that the minimal basis of
gauge theory amplitudes with $n$ external legs grows only like
$(n-3)!$~\cite{BjerrumBohr:2009rd}, and not like $(n-2)!$ as
would be concluded on the basis of the field theory
Kleiss-Kuijf relations~\cite{Kleiss:1988ne,DelDuca:1999rs}
alone. Quite recently both sets of relations have been proven
by means of BCFW-recursion~\cite{Feng:2010my} (see
also~\cite{Jia:2010nz}), thus also there circumventing the need
to go through string theory first.

The fact that BCJ-relations are central to KLT-relations and to
the various rewritings of them becomes quite obvious in this
paper. In this connection it is interesting that the first
example of a BCJ-like relation is implicitly contained in
ref.~\cite{Kawai:1985xq} by noting that alternative
factorizations of closed string amplitudes into products of
open string amplitudes yield different expressions. Equating
the two, a new identity is inferred, and this is indeed one
particular BCJ-relation. From our perspective today, this is
not surprising since it is
monodromy~\cite{BjerrumBohr:2009rd,Stieberger:2009hq} that is
the mechanism behind the stringy BCJ-relations. However, our
aim in this paper is to phrase the whole discussion in terms of
quantum field theory concepts alone. It turns out that the one
surviving property from string theory that we need to use is
precisely just the information encoded in the field theory
BCJ-relations.

The particular versions of the field theory KLT-relations that
were proven in ref.~\cite{BjerrumBohr:2010ta} are not those
commonly referred to before and which were conjectured in
ref.~\cite{Bern:1998sv}. The relations proven
in~\cite{BjerrumBohr:2010ta} keep two external legs fixed while
summing over permutations of the remaining. This is ideally
suited for a proof based on on-shell recursion. The
disadvantage is, as will be reviewed below, that they use a
representation that contains an apparent singularity when all
legs are on-shell. They therefore need to be supplemented with
a suitable regularization. Once regulated, the on-shell field
theory BCJ-relations can be used to remove the apparent
singularity in a systematic manner. In this way, one ends up
with relations that can be written in more familiar forms.
While this is sufficient as a matter of principle, one would
still like to have a corresponding direct proof of the
conjectured KLT-formula for arbitrary $n$-point amplitudes that
can be found in ref.~\cite{Bern:1998sv}. In the course of
another investigation dealing with a new set of gauge theory
identities~\cite{BjerrumBohr:2010zb}\footnote{For the
generalization to the full ${\cal N}=4$ multiplet,
see~\cite{Feng:2010br}. An interpretation of these identities
in terms of vanishing gravity amplitudes with complex scalars
has recently been given by Tye and Zhang~\cite{Tye:2010kg}.} we
uncovered a much wider sequence of explicit field theory
KLT-relations that are all, via BCJ-relations,
equivalent.\footnote{After the completion
of~\cite{BjerrumBohr:2010zb} we discovered another
paper~\cite{Abe:2005se} where an arbitrariness of the
expression conjectured in~\cite{Bern:1998sv} was noted. We
believe, but have not checked in detail, that this
arbitrariness is the same that we exploited
in~\cite{BjerrumBohr:2010zb} and which we will prove explicitly
below.} The purpose of this paper is to prove these explicit
field theory KLT-relations.

The plan of our paper is as follows. In section 2 we introduce
the different versions of field theory KLT-relations we will
prove by BCFW-recursion. We also discuss two very useful
functions that form the backbone of KLT-relations in the field
theory limit. It is due to the remarkable properties of these
functions that we can so easily move between different
formulations. In particular, these functions encode the
BCJ-relations and almost miraculously cancel poles so as to
provide gravity amplitudes from products of two gauge theory
amplitudes. In section 3 we give the explicit proofs. It is
interesting that these proofs simplify greatly by having at our
disposal several different versions of the KLT-relations. Since
these different versions are related through BCJ-relations this
shows how important these BCJ-relations are for establishing
the field theory proof of KLT-relations. Section 4 contains our
conclusions.

\section{KLT-relations}

We begin by presenting several different forms of the
KLT-relations in the field theory limit. The first is the
version presented and proven in ref.~\cite{BjerrumBohr:2010ta}.
It keeps just two legs fixed in each gauge theory amplitude and
sums over $(n-2)!$ permutations. This expression has the
advantage of having a high degree of manifest permutation
symmetry, but the price one pays is that it contains,
necessarily, many terms. As mentioned in the Introduction, it
also has an apparent singularity and therefore requires
regularization. Once regularized, one sees that the apparent
pole is canceled by a corresponding term in the numerator. This
KLT-form in which just two legs are kept fixed (rather than
three, see below) is ideally suited for a proof using on-shell
recursion in the form of BCFW-shifts, the two legs that are
kept fixed naturally lending themselves to the needed momentum
shifts.

Next we introduce a whole family of alternative KLT-relations
that have less manifest permutation symmetry. They contain
fewer terms and have no need of a regularization. We have
already noted these alternative forms in a paper on new
identities in gauge theories~\cite{BjerrumBohr:2010zb}. Here we
will explicitly prove that they are all equivalent. Because
they contain fewer terms, they might be regarded as more useful
in practice. Among these new forms is also the originally
conjectured $n$-point KLT-relation  of~\cite{Bern:1998sv}. It
is thus a special case.

Although these more compact KLT-expressions are not ideally
suited for a proof based on BCFW-recursion with just two legs
shifted, we can nevertheless carry such a proof through. This
will be shown below.

\subsection{Two useful functions}

Before presenting the KLT-relations let us for the sake of
completeness introduce some useful definitions and the notation
that will be used throughout the paper.

We begin with the following
$\mathcal{S}$-function~\cite{BjerrumBohr:2010ta}
\bea {\cal S}[i_1,\ldots,i_k|j_1,\ldots,j_k] & = &
\prod_{t=1}^k \big(s_{i_t 1}+\sum_{q>t}^k \theta(i_t,i_q)
s_{i_t i_q}\big)\,,~~~\label{S-def} \eea
were $\theta(i_t, i_q)$ is zero when the pair $(i_t,i_q)$ has
same ordering at both sets $\mathcal{I}\equiv \{
i_1,\ldots,i_k\}$ and $\mathcal{J}\equiv \{j_1,\ldots,j_k\}$,
but otherwise it is unity (we will everywhere use
$s_{ij}\equiv(p_i+p_j)^2$ and $s_{i\ldots
k}\equiv(p_i+\ldots+p_k)^2$). The function ${\cal S}$ is
nothing but a rewriting of the $f$-function defined
in~\cite{Bern:1998sv}, now with a notation that we find more
compelling. In this definition, the set ${\cal J}$ is the
reference ordering set, $i.e.$, this set provides the standard
ordering. The set ${\cal I}$ determines the given factor in the
product by comparing with the set ${\cal J}$. It might be
useful to illustrate the definition with a few examples
\bean {\cal S}[2,3,4|2,4,3] & = & s_{21} (s_{31}+s_{34})
s_{41}\,,~~~ {\cal
S}[2,3,4|4,3,2]=(s_{21}+s_{23}+s_{24})(s_{31}+s_{34}) s_{41}\,.
\eean
Note that for each leg $i$ there is a term $s_{i 1}$. In other
words, the momentum $p_1$ has been singled out and it plays a
special role. When we have several functions ${\cal S}$ with
different choices of $p_1$, we will write ${\cal S}[{\cal
I}|{\cal J}]_{p_1}$ to avoid confusion.

One important property of the function ${\cal S}$ is
the reflection symmetry
\bea {\cal S}[i_1,\ldots,i_k|j_1,\ldots,j_k]={\cal
S}[j_k,\ldots,j_1|i_k,\ldots,i_1]\,,~~~\label{S-rev} \eea
where we have exchanged the two sets and at the same time
reversed the ordering of each set. The reason for this identity
is clear. First we notice that if $(x,y)$ at both sets ${\cal
I}$ and ${\cal J}$ have the same (different) ordering, they
have same (different) ordering after swapping the sets and
reversing the order. If the ordering is the same for the pair
$(x,y)$ in both sets there is no corresponding $s_{xy}$. If the
ordering is different for the pair $(x,y)$ we have, before
swapping, ${\cal S}[ A,x,B,y,C|D,y,E,x,F]$. We thus have a
factor $(s_{x1}+s_{xy}+\ldots)$, while after the exchange we
have ${\cal S}[ F,x,E,y,D|C,y,B,x,A]$ and hence the same factor
$(s_{x1}+s_{xy}+\ldots)$.

It is also convenient to introduce a dual ${\cal \W
S}$-function
\begin{align}
\widetilde{\mathcal{S}}[i_1,\ldots,i_{k}|j_1,\ldots,j_{k}] \equiv
\prod_{t=1}^{k} \big(s_{j_t n}+\sum_{q<t} \theta(j_q,j_t)
s_{j_t j_q}\big) \label{dualSdef}\,,
\end{align}
where again $\theta(j_a,j_b)$ is zero if $j_a$ sequentially
comes before $j_b$ in $\{i_1,\ldots,i_{k}\}$, and otherwise it
is unity. This dual form matches precisely the
$\bar{f}$-function defined in~\cite{Bern:1998sv}, again
rewritten in a way that we find convenient. For this dual
${\cal \W S}$, we will think of the standard ordering as given
by the set ${\cal I}$. It is worth emphasizing that the
summation in (\ref{dualSdef}) is over $q<t$, $i.e.$, over
elements to the left of $j_t$ in the set ${\cal J}$. As an
example,
\bean {\cal \W S}[2,3,4|4,3,2]_{p_5}= s_{45} (s_{35}+s_{34})
(s_{25}+s_{23}+s_{24})\,. \eean
And $\widetilde{\mathcal{S}}$ has the same reflection symmetry
as $\mathcal{S}$
\bea {\cal \W S}[i_1,\ldots,i_{k}|j_1,\ldots,j_{k}]={\cal \W
S}[j_{k},\ldots,j_1|i_{k},\ldots,i_1]\,.~~~\label{dual-S-rev}
\eea

The functions ${\cal S}$ and ${\cal \W S}$ have several other
useful properties. Considered as `operators' they annihilate
amplitudes in the following sense
\begin{align}
 \sum_{\a\in S_{n-2}}{\cal
S}[\a_{2,n-1}|j_2,\ldots,j_{n-1}]_{p_1}
A_n(n,\a_{2,n-1},1) = 0\,.~~~\label{I-com}
\end{align}
We will throughout use the shorthand notation: $\a_{2,n-1}$ for
the ordering $\a$ of legs $2,3,\ldots,n-1$ in amplitudes. This
property is in fact nothing but a rephrasing of BCJ-relations.
To see this, let us first consider a six-point example with
${\cal J}=(2,3,4,5)$. The sum of permutation $\a_{2,5}$ can be
divided into permutations of the set $\{ 2,3,4 \}$ plus $5$
placed at all possible positions. Considering one particular
ordering of $\{ 2,3,4 \}$, for example the ordering $(3,4,2)$,
we have the sum
\begin{align}
&  {\cal S}[ 3,4,2,5 |2,3,4,5]
A_6(6,3,4,2,5,1) + {\cal S}[ 3,4,5,2 |2,3,4,5] A_6(6,3,4,5,2,1)\nn
& + {\cal S}[ 3,5,4,2 |2,3,4,5] A_6(6,3,5,4,2,1)+ {\cal S}[5,
3,4,2 |2,3,4,5] A_6(6,5,3,4,2,1)\nn
 &=  (s_{31}+s_{32}) (s_{41}+s_{42})s_{21}\big[ s_{51}
A_6(6,3,4,2,5,1) + (s_{51}+s_{52}) A_6(6,3,4,5,2,1)\nn
& +
(s_{51}+s_{52}+s_{54})A_6(6,3,5,4,2,1)+
(s_{51}+s_{52}+s_{54}+s_{53})A_6(6,5,3,4,2,1)\big]\,,
\end{align}
which indeed vanishes as a consequence of the BCJ-relation
\begin{align}
0 ={}& s_{51} A_6(6,3,4,2,5,1) + (s_{51}+s_{52}) A_6(6,3,4,5,2,1) \nonumber \\
& + (s_{51}+s_{52}+s_{54})A_6(6,3,5,4,2,1)+
(s_{51}+s_{52}+s_{54}+s_{53})A_6(6,5,3,4,2,1)\,.
\label{BCJ6pt}
\end{align}
The above relation, \textit{i.e.}~eq.~\eqref{BCJ6pt}, is given
in the form that we from now on will denote as a
\textit{fundamental} BCJ-relation. Note how a leg $j$ (above
$j=5$) moves one leg to the left in each term and picks up an
additional factor of $s_{ji}$ (above $s_{5i}$), where $i$ is
the leg just passed.

With the above example at hand it is easy to see how this
argument generalizes. We divide the sum of $\a_{2,n-1}$ into
the sum of groups where all, except $j_{n-1}$, have fixed
ordering, and then insert $j_{n-1}$ at any place. For each
group all factors from $\mathcal{S}$ are the same except the
factor contributing from $j_{n-1}$. Like in the above example
this will provide us with a fundamental BCJ-relation and thus
vanish.

There is clearly an analogous relation for
$\widetilde{\mathcal{S}}$:
\begin{align}
\sum_{\gamma\in S_{n-2}} {\cal \W S}[i_2,\ldots,i_{n-1}|\gamma_{2,n-1}]_{p_{n}} \W
A_n(n,\gamma_{2,n-1}, 1) = 0\,.
\end{align}
%

\subsection{New KLT forms}

We are now ready to present the actual KLT-relations. The new
form that was presented and proven in~\cite{BjerrumBohr:2010ta}
is
\bea M_n=(-1)^n\sum_{\gamma,\b}{\W A_n(n,\gamma_{2,n-1},1)
{\cal S}[ \gamma_{2,n-1}|\b_{2,n-1}]_{p_1}
A_n(1,\b_{2,n-1},n)\over s_{12\ldots(n-1)}}\,.~~~\label{newKLT}
\eea
It also has a dual form given by
\bea M_n  =  (-1)^n\sum_{\b,\gamma} {A_n (1,\b_{2,n-1},n) {\cal
\W S}[\b_{2,n-1}|\gamma_{2,n-1}]_{p_{n}} \W
A_n(n,\gamma_{2,n-1}, 1)\over s_{23\ldots
n}}\,,~~~\label{New-KLT-dual} \eea
where $A_n$ and $\widetilde{A}_n$ are gauge theory amplitudes
and $M_n$ is the gravity amplitude. Although only
eq.~\eqref{newKLT} was proven in~\cite{BjerrumBohr:2010ta},
eq.~\eqref{New-KLT-dual} can obviously be proven in exactly the
same manner and we will not repeat the proof here. As already
discussed in~\cite{BjerrumBohr:2010ta}, the above forms are
singular when all legs are on-shell. They require
regularization, which fortunately is easily implemented, see
ref.~\cite{BjerrumBohr:2010ta} for details.

\subsection{The general KLT expression}

In terms of the $\mathcal{S}$-functions defined above the
explicit $n$-point KLT-relation conjectured
in~\cite{Bern:1998sv} can be written as
\begin{align}
M_n = & (-1)^{n+1}\!\!\! \sum_{\sigma\in S_{n-3}}\sum_{\a\in
S_{j-1}}\sum_{\b\in S_{n-2-j}}
A_n(1,\sigma_{2,j},\sigma_{j+1,n-2},
n-1,n) {\cal S}[
\a_{\sigma(2),\sigma(j)}|\sigma_{2,j}]_{p_1}\nn &
\times {\cal \W S}[\sigma_{j+1,n-2}
|\b_{\sigma(j+1),\sigma(n-2)}]_{p_{n-1}} \W
A_n(\a_{\sigma(2),\sigma(j)},1,n-1,
\b_{\sigma(j+1),\sigma(n-2)},n)\,,\label{KLT-bern}
\end{align}
where $j=[n/2]$ is a fixed number ($[x]$ is the integer value
of $x$). However, by the use of BCJ-relations we will see that
the formula (\ref{KLT-bern}) actually holds for arbitrary $j$.
It therefore presents us with a whole family of
equivalent KLT-relations~\cite{BjerrumBohr:2010zb}.

To see this generalization, let us first start with an example
of $n=8$ before moving to the general $n$-point case. The term
multiplying $A_8(1,(2,3,4),(5,6),7,8)$ is given by
\bea
& & \sum_{\a\in S_3}\sum_{\b\in S_{2}}
 {\cal S}[
\a_{2,4}|2,3,4]_{p_1} {\cal \W S}[5,6 |\b_{5,6}]_{p_7}
\widetilde{A}_8(\a_{2,4},1,7,\b_{5,6},8)\,. \eea
The sum of  permutations $\a_{2,4}$ can be divided into groups
of fixed permutation of $\W\a_{2,3}$ plus all possible
insertions of $4$. Thus, with fixed permutation $\b$, for
example $\b_{5,6}=(6,5)$, and fixed permutation $\W\a$, for
example $\W\a_{2,3}=(3,2)$, the sum is given by
\bea && \hspace{0cm}\ \ \ \,\label{8ptmidcal}
 {\cal S}[
3,2,4|2,3,4]_{p_1} {\cal \W S}[5,6 |6,5]_{p_7} \W
A_8(3,2,4,1,7,6,5,8)\nn && \hspace{0cm}+{\cal S}[
3,4,2|2,3,4]_{p_1} {\cal \W S}[5,6 |6,5]_{p_7} \W
A_8(3,4,2,1,7,6,5,8)\nn
  &&\hspace{0cm}+  {\cal S}[
4,3,2|2,3,4]_{p_1} {\cal \W S}[5,6 |6,5]_{p_7} \W
A_8(4,3,2,1,7,6,5,8)\nn &&\hspace{0cm} =  s_{67} (s_{57}+
s_{56}) (s_{31}+s_{32})s_{21} \left[ s_{41}\W
A_8(3,2,4,1,7,6,5,8)\right.\\
 &&\hspace{0cm}+\left. (s_{41}+s_{42})\W
A_8(3,4,2,1,7,6,5,8)\right. \left.+ (s_{41}+s_{42}+s_{43})\W
A_8(4,3,2,1,7,6,5,8)\right].\nonumber  \eea
We now use the fundamental BCJ-relation
\begin{align}
0 ={}& s_{41} \widetilde{A}_8(3,2,4,1,7,6,5,8) + (s_{41}+s_{42})\widetilde{A}_8(3,4,2,1,7,6,5,8)
\nn &+(s_{41}+s_{42}+s_{43})\widetilde{A}_8(4,3,2,1,7,6,5,8)
+ (s_{41}+s_{42}+s_{43}+s_{48})\widetilde{A}_8(3,2,1,7,6,5,4,8)
\nn &+ (s_{41}+s_{42}+s_{43}+s_{48}+s_{45})\widetilde{A}_8(3,2,1,7,6,4,5,8)
\nn &+ (s_{41}+s_{42}+s_{43}+s_{48}+s_{45}+s_{46})\widetilde{A}_8(3,2,1,7,4,6,5,8)\,,
\end{align}
where 4 is the leg being moved to the left. However, by momentum conservation we also have
\begin{eqnarray}
& s_{41}+s_{42}+s_{43}+s_{48} = -(s_{47}+ s_{45}+s_{46}),  \qquad
s_{41}+s_{42}+s_{43}+s_{48}+s_{45} = - (s_{47}+ s_{46}), & \nn
&s_{41}+s_{42}+s_{43}+s_{48}+s_{45}+s_{46} = -s_{47}\,,&
\end{eqnarray}
and eq.~\eqref{8ptmidcal} takes the form
\bea &&\!\!\!\!s_{67} (s_{57}+ s_{56}) (s_{31}+s_{32})s_{21}
\left[ (s_{47}+ s_{45}+s_{46})  \W
A_8(3,2,1,7,6,5,4,8)\right.\nn &&\left.\hspace{4.6cm}+ (s_{47}+
s_{46}) \W A_8(3,2,1,7,6,4,5,8)\right. \left.+s_{47} \W
A_8(3,2,1,7,4,6,5,8)\right]\nn&&\!\!\!\!  =\!  {\cal
S}[3,2|2,3]_{p_1}\!\!\left[{\cal \W S}[4,5,6|6,5,4]_{p_7}\W
A_8(3,2,1,7,6,5,4,8)\!+\!{\cal \W S}[4,5,6|6,4,5]_{p_7}\W
A_8(3,2,1,7,6,4,5,8) \right.\nn && \left. +{\cal \W
S}[4,5,6|4,6,5]_{p_7} \W A_8(3,2,1,7,4,6,5,8) \right]. \eea
Summing up all possible permutations $\W\a,\b$ we find that the
term multiplying
$$
A_8(1,(2,3,4),(5,6),7,8) = A_8(1,(2,3),(4,5,6),7,8)
$$
is given by
\bea
& & \sum_{\a\in S_2}\sum_{\b\in S_{3}}
 {\cal S}[
\a_{2,3}|2,3]_{p_1} {\cal \W S}[4,5,6 |\b_{4,6}]_{p_7} \W
A_8(\a_{2,3},1,7,\b_{4,6},8)\,, \eea
where leg $4$ has been moved from the set of ${\cal S}$ into the
set of ${\cal \W S}$. This can be done for any of the $\sigma$
permutations and thereby give us the eight-point relation in
the form of eq.~\eqref{KLT-bern} but with $j$ shifted down by
one. Repeating this argument we can get to any of the $j$-value
forms we want.

We can now follow the same procedure for general $n$ and obtain
the following relation
\begin{align}
&\sum_{\alpha,\beta} \mathcal{S}[\alpha_{i_2,i_j}|i_2,\ldots,i_j]_{p_1}
\widetilde{\mathcal{S}}[i_{j+1},\ldots,i_{n-2}|\beta_{i_{j+1},i_{n-2}}]_{p_{n-1}}
\widetilde{A}_n(\alpha_{i_2,i_j},1,n-1,\beta_{i_{j+1},i_{n-2}},n) \nonumber \\
&=\sum_{\alpha',\beta'} \mathcal{S}[\alpha'_{i_2,i_{j-1}}|i_2,\ldots,i_{j-1}]_{p_1}
\widetilde{\mathcal{S}}[i_j,i_{j+1},\ldots,i_{n-2}|\beta'_{i_j,i_{n-2}}]
\widetilde{A}_n(\alpha'_{i_2,i_{j-1}},1,n-1,\beta'_{i_j,i_{n-2}},n)\,,
\label{jrel}
\end{align}
from which the equivalence of eq.~\eqref{KLT-bern} for any
$j$-value follows. The general proof of eq.~\eqref{jrel} is
relegated to Appendix \ref{appA}.

One interesting application of this result is that we can shift
$j$ all the way to make the left- or right-hand part empty,
\textit{i.e.} we can choose $j=1$ or $j=n-2$. Then we get the
following two expressions~\cite{BjerrumBohr:2010zb}
\begin{align}
M_n = (-1)^{n+1}\sum_{\sigma,\widetilde{\sigma} \in
S_{n-3}} \widetilde{A}_n(n-1,n,\widetilde{\sigma}_{2,n-2},1)
\mathcal{S}[\widetilde{\sigma}_{2,n-2}|\sigma_{2,n-2}]_{p_1}
A_n(1,\sigma_{2,n-2},n-1,n)\,,
\label{pureKLT}
\end{align}
and
\begin{align}
M_n = (-1)^{n+1}\sum_{\sigma,\widetilde{\sigma} \in S_{n-3}}
A_n(1,\sigma_{2,n-2},n-1,n)
\widetilde{\mathcal{S}}[\sigma_{2,n-2}|\widetilde{\sigma}_{2,n-2}]_{p_{n-1}}
\widetilde{A}_n(1,n-1,\widetilde{\sigma}_{2,n-2},n)\,,
\label{dualKLT}
\end{align}
which are exactly related to each other by
\begin{align}
&\!\sum_{\widetilde{\sigma}}\!\widetilde{A}_n(n-1,n,\widetilde{\sigma}_{2,n-2},1)
\mathcal{S}[\widetilde{\sigma}_{2,n-2}|\sigma_{2,n-2}]_{p_1}
\!=\! \sum_{\widetilde{\sigma}}\! \widetilde{A}_n(1,n-1,\widetilde{\sigma}_{2,n-2},n)
\widetilde{\mathcal{S}}[\sigma_{2,n-2}|\widetilde{\sigma}_{2,n-2}]_{p_{n-1}},
\label{obs}
\end{align}
following from repeated use of eq.~\eqref{jrel}.

We will call the latter form, \textit{i.e.}
eq.~\eqref{dualKLT}, the {\em dual form} of (\ref{pureKLT}). It is
important to stress again that all these forms are completely
equivalent. As we have explained above and prove in detail
in Appendix A, they are related to each
other by use of BCJ-relations through eq.~\eqref{jrel}.

\section{Field theory proof of KLT-relations}

We now turn to the field theory proof of the whole family of
KLT-expressions introduced above. As we will see, the $(n-2)!$
symmetric form will be an important ingredient even for the
proof of the relations that are only manifestly $(n-3)!$
symmetric.

Since all forms in the family of eq.~\eqref{KLT-bern}
(\textit{i.e.} with arbitrary $j$) are related to each other by
BCJ-relations, we are free to choose any of the versions we
find most convenient. Proving one version clearly proves them
all.

As in~\cite{BjerrumBohr:2010ta} we will give an induction proof
by means of BCFW-recursion. Specifically, we will make a
$(1,n-1)$ BCFW-shift and consider the integral
\begin{align}
0 = \oint \frac{dz}{z}M_n(z) = M_n(0) + \sum (\mathrm{residues}\:\:\mathrm{for}\:\: z\neq 0).
\end{align}
Evidently, if there are boundary terms to this contour
integral, they are not included here. It has been shown in
refs.~\cite{zscaling,zscaling2} that the fall-off at $z \to
\infty$ of the graviton amplitude $M_n(z)$ is even stronger
than one could naively have guessed, and goes as $1/z^2$. Our
aim is to show that the sum of residues exactly gives us the
BCFW-expansion of the $n$-point gravity amplitude through use
of only lower-point KLT relations. It is easy to see that the
relations are indeed satisfied for low-point relations.

To simplify the expressions as much as possible we will mainly
use the simple versions given by eq.~\eqref{pureKLT} and
\eqref{dualKLT}. For convenience we discard the overall sign
$(-1)^{n+1}$, which is readily reinstated into the proof.

We start by considering all the residues coming from poles of
the form $s_{12\ldots k}$ (we can compute the residue of the
pole $s_{\widehat{1}2\ldots k}$ as $\ \sim\lim_{z\rightarrow
z_{12\ldots k}} \big[ s_{\widehat{1}2\ldots k}(z) M_n(z)
\big]/s_{12\ldots k}$, where $z_{12\ldots k}$ is the $z$-value
that makes $s_{\widehat{1}2\ldots k}$ go on-shell). \\ \\ There
are three cases
\begin{itemize}
 \item (A-1) The pole appears only in $\widetilde{A}_n$.
 \item (A-2) The pole appears only in $A_n$.
 \item (B) The pole appears both in $\widetilde{A}_n$ and
     $A_n$.
\end{itemize}

Starting with (A-1) we see that the contributing terms from
eq.~\eqref{pureKLT} will be of the form
\begin{align}
&\sum_{\sigma,\widetilde{\sigma},\alpha}
\widetilde{A}_n(\widehat{n-1},n,\widetilde{\sigma}_{k+1,n-2},\alpha_{2,k},
\widehat{1})
\mathcal{S}[\widetilde{\sigma}_{k+1,n-2}\alpha_{2,k}
|\sigma_{2,n-2}]_{\widehat{p}_1}
A_n(\widehat{1},\sigma_{2,n-2},\widehat{n-1},n)\,,
\end{align}
and we therefore get the residue
\begin{align}
&\sum_{\sigma,\widetilde{\sigma},\alpha} \frac{\sum_h\widetilde{A}(
\widehat{n-1},n,\widetilde{\sigma}_{k+1,n-2},\widehat{P}^h)
\widetilde{A}(-\widehat{P}^{-h},\alpha_{2,k},\widehat{1})}{s_{12\ldots k}}
\mathcal{S}[\widetilde{\sigma}_{k+1,n-2}\alpha_{2,k}
|\sigma_{2,n-2}]_{\widehat{p}_1} \nonumber \\
&\hspace{10.3cm} \times  A_n(\widehat{1},\sigma_{2,n-2},\widehat{n-1},n)\,.
\end{align}
We now use the important factorization property (which follows
from the definition of $\mathcal{S}$)
\begin{align}
\mathcal{S}[\widetilde{\sigma}_{k+1,n-2}\alpha_{2,k}
|\sigma_{2,n-2}]_{\widehat{p}_1}
= \mathcal{S}[\alpha_{2,k}|\rho_{2,k}]_{\widehat{p}_1}
\times (\mathrm{a}\,\, \mathrm{factor}\,\,
\mathrm{independent}\,\, \mathrm{of}\, \alpha)\,,
\end{align}
where $\rho$ denotes the relative ordering of leg
$2,3,\ldots,k$ in the set $\sigma$. It follows that this contribution
contains an expression like
\begin{align}
\sim \sum_{\alpha} \widetilde{A}(-\widehat{P}^{-h},
\alpha_{2,k},\widehat{1}) \mathcal{S}[\alpha_{2,k}
|\rho_{2,k}]_{\widehat{p}_1} = 0\,,
\end{align}
which, as we have seen above, vanishes due to BCJ-relations. We
therefore conclude that contributions from (A-1) vanish
altogether.

With a similar argument we note that contributions from (A-2)
also vanish.

Moving on to (B) the contributing terms in
eq.~\eqref{pureKLT} take the form
\begin{align}
&\sum_{\sigma,\widetilde{\sigma},\alpha,\beta}
\widetilde{A}_n(\widehat{n-1},n,\widetilde{\sigma}_{k+1,n-2},
\alpha_{2,k},\widehat{1})
\mathcal{S}[\widetilde{\sigma}_{k+1,n-2}\alpha_{2,k}
|\beta_{2,k}\sigma_{k+1,n-2}]_{\widehat{p}_1} \nonumber \\
&\hspace{9.5cm} \times A_n(\widehat{1},\beta_{2,k},
\sigma_{k+1,n-2},\widehat{n-1},n)\,.
\label{midB}
\end{align}
Using the factorization property
\begin{align}
\mathcal{S}[\widetilde{\sigma}_{k+1,n-2}\alpha_{2,k}
|\beta_{2,k}\sigma_{k+1,n-2}]_{\widehat{p}_1}
= \mathcal{S}[\alpha_{2,k}|\beta_{2,k}]_{\widehat{p}_1} \times
\mathcal{S}[\widetilde{\sigma}_{k+1,n-2}|\sigma_{k+1,n-2}]_{\widehat{P}}\,,
\end{align}
the residue for $s_{12\ldots k}$ can be written as
\begin{align}
& \frac{1}{s_{12\ldots k}}\sum_h\sum_{\alpha,\beta}
\frac{\widetilde{A}(-\widehat{P}^h,\alpha_{2,k},\widehat{1})
\mathcal{S}[\alpha_{2,k}|\beta_{2,k}]_{\widehat{p}_1}
A(\widehat{1},\beta_{2,k},-\widehat{P}^h)}{s_{\widehat{1}2\ldots k}} \nonumber \\
&\times \sum_{\sigma,\widetilde{\sigma}}
\widetilde{A}(\widehat{n-1},n,\widetilde{\sigma}_{k+1,n-2},\widehat{P}^{-h})
\mathcal{S}[\widetilde{\sigma}_{k+1,n-2}|\sigma_{k+1,n-2}]_{\widehat{P}}
A(\widehat{P}^{-h},\sigma_{k+1,n-2},\widehat{n-1},n)\,,
\end{align}
where we have used the vanishing of all the mixed-helicity
contributions~\cite{BjerrumBohr:2010ta,BjerrumBohr:2010zb}.

Remarkably, the sum over $\alpha$ and $\beta$ is precisely the
singular KLT-form of ref.~\cite{BjerrumBohr:2010ta}. As shown
in that paper, it is equal to
$M_{k+1}(\widehat{1},2,\ldots,k,-\widehat{P}^h)$. Moreover, the
sum over $\sigma$ and $\widetilde{\sigma}$ is an $n-k+1$ point
version of eq.~\eqref{pureKLT} and hence, by induction, equal
to $M_{n-k+1}(k+1,\ldots,\widehat{P}^{-h})$, \textit{i.e.}
\begin{align}
\sum_h \frac{M_{k+1}(\widehat{1},2,\ldots,k,-\widehat{P}^h)
M_{n-k+1}(k+1,\ldots,\widehat{P}^{-h})}{s_{12\ldots k}}\,.
\end{align}
{}From this we conclude that we get the correct BCFW-contributions
to the $n$-point gravity amplitude for all poles
of the $s_{12\ldots k}$ kind, and by the manifest $(n-3)!$ symmetry
also for all poles related to these by permutation of leg
$2,3,\ldots,n-2$.

However, this is not the end of the proof. Because of the only
$(n-3)!$ manifest symmetry we also need to explicitly consider
pole contributions involving leg 1 and $n$, \textit{i.e.} poles
of the form $s_{12\ldots n \ldots k} = s_{k+1\ldots n-1}$. To
investigate these contributions we will use the dual form,
eq.~\eqref{dualKLT}. This is allowed since we have seen that
eq.~\eqref{pureKLT} and~\eqref{dualKLT} are just two different
ways of writing the same quantity.

Again we start with (A-1), but this time considering the
$s_{12\ldots n \ldots k}$ pole and using eq.~\eqref{dualKLT}
instead of the equivalent eq.~\eqref{pureKLT}. The residue
takes the form
\begin{align}
&\sum_{\sigma,\widetilde{\sigma},\alpha}
A_n(\widehat{1},\sigma_{2,n-2},\widehat{n-1},n)
\widetilde{\mathcal{S}}[\sigma_{2,n-2}
|\widetilde{\sigma}_{k+1,n-2}\alpha_{2,k}]_{\widehat{p}_{n-1}} \nonumber \\
&\hspace{6cm} \times \sum_h \frac{\widetilde{A}(\widehat{n-1},
\widetilde{\sigma}_{k+1,n-2},\widehat{P}^h)
\widetilde{A}(-\widehat{P}^{-h},\alpha_{2,k},n,\widehat{1})}{s_{12\ldots n\ldots k}}\,.
\end{align}
Using the factorization property
\begin{align}
\widetilde{\mathcal{S}}[\sigma_{2,n-2}
|\widetilde{\sigma}_{k+1,n-2}\alpha_{2,k}]_{\widehat{p}_{n-1}}
={}& \widetilde{\mathcal{S}}[\rho_{k+1,n-2}
|\widetilde{\sigma}_{k+1,n-2}]_{\widehat{p}_{n-1}}
&\!\!\!\times (\mathrm{a}\,\, \mathrm{factor}\,\,
\mathrm{independent}\,\, \mathrm{of}\, \widetilde{\sigma})\,,
\end{align}
where $\rho$ denotes the relative ordering of leg
$k+1,\ldots,n-2$ in set $\sigma$, we once again see that the
contribution contains a factor of
\begin{align}
\sim \sum_{\widetilde{\sigma}} \widetilde{A}(\widehat{n-1},
\widetilde{\sigma}_{k+1,n-2},\widehat{P}^h)
\widetilde{\mathcal{S}}[\rho_{k+1,n-2}
|\widetilde{\sigma}_{k+1,n-2}]_{\widehat{p}_{n-1}} = 0\,,
\end{align}
that vanishes due to BCJ-relations.

Similar arguments apply to (A-2) for the $s_{12\ldots n\ldots
k}$ pole.

Now we consider (B) for a $s_{12\ldots n\ldots k}$ pole. In
this case the contributing terms of eq.~\eqref{dualKLT} have
the form
\begin{align}
&\sum_{\sigma,\widetilde{\sigma},\alpha,\beta}
A_n(\widehat{1},\beta_{2,k},\sigma_{k+1,n-2},\widehat{n-1},n)
\widetilde{\mathcal{S}}[\beta_{2,k}\sigma_{k+1,n-2}
|\widetilde{\sigma}_{k+1,n-2}\alpha_{2,k}]_{\widehat{p}_{n-1}} \nonumber \\
& \hspace{9cm} \times \widetilde{A}_n(\widehat{1},
\widehat{n-1},\widetilde{\sigma}_{k+1,n-2},\alpha_{2,k},n)\,,
\end{align}
and $\widetilde{\mathcal{S}}$ satisfies the factorization
property
\begin{align}
\widetilde{\mathcal{S}}[\beta_{2,k}\sigma_{k+1,n-2}
|\widetilde{\sigma}_{k+1,n-2}\alpha_{2,k}]_{\widehat{p}_{n-1}}
={}& \widetilde{\mathcal{S}}[\sigma_{k+1,n-2}
|\widetilde{\sigma}_{k+1,n-2}]_{\widehat{p}_{n-1}}\widetilde{\mathcal{S}}[\beta_{2,k}
|\alpha_{2,k}]_{\widehat{P}}\,.
\end{align}
Hence the residue can be written
\begin{align}
&\frac{1}{s_{12\ldots n\ldots k}}\sum_h \sum_{\alpha,\beta}
A(\widehat{1},\beta_{2,k},\widehat{P}^h,n)
\widetilde{\mathcal{S}}[\beta_{2,k}|\alpha_{2,k}]_{\widehat{P}}
\widetilde{A}(\widehat{1},\widehat{P}^h,\alpha_{2,k},n) \nonumber \\
&\times \sum_{\widetilde{\sigma},\sigma}
\frac{A(-\widehat{P}^{-h},\sigma_{k+1,n-2},\widehat{n-1})
\widetilde{\mathcal{S}}[\sigma_{k+1,n-2}
|\widetilde{\sigma}_{k+1,n-2}]_{\widehat{p}_{n-1}}
\widetilde{A}(\widehat{n-1},\widetilde{\sigma}_{k+1,n-2},
-\widehat{P}^{-h})}{s_{k+1\ldots \widehat{n-1}}}\,,
\end{align}
where we have used $s_{\widehat{1}2\ldots n\ldots
k}=s_{k+1\ldots\widehat{n-1}}$, and the vanishing of
mixed-helicity contributions. We see that the first term is
just a lower-point version of eq.~\eqref{dualKLT}, and the
second term the singular dual KLT-form, \textit{i.e.}
\begin{align}
\sum_h \frac{M_{k+2}(\widehat{1},2,\ldots,n,
\widehat{P}^h)M_{n-k}(k+1,\ldots,\widehat{n-1},
-\widehat{P}^{-h})}{s_{12\ldots n\ldots k}}\,.
\end{align}
We hereby see that we once again obtain the correct BCFW
expansion for all $s_{12\ldots n\ldots k}$ poles, and by the
symmetry of leg $2,\ldots,n-2$ also for all poles related to
these by a permutation.

The above analysis covers all residues for the $n$-point
KLT-relation, which we see give the full BCFW-expansion for the
$n$-point gravity amplitude.  We stress that we have
\textit{not} assumed any permutation symmetries beyond that
which are already manifest in the expressions. This therefore
also constitutes a proof that these KLT-expression are fully
symmetric under permutations since the gravity amplitudes have
this property. It would be most tedious to prove this fact by
more direct means.

\section{Conclusion}

We have provided a field theory proof by induction of a whole
new class of KLT-expressions that are all equivalent to the
particular case conjectured in ref.~\cite{Bern:1998sv}. Only
general properties of the $S$-matrix in quantum field
theory~\cite{Paolo:2007} have been assumed.

As a byproduct of our study, we have demonstrated the validity
of various rewritings of KLT-relations that we think will be of
use. The two forms given in (\ref{pureKLT}) and (\ref{dualKLT})
and which have manifest $(n-3)!$ permutation symmetry are of
particular interest. These two expressions are very similar to
the new KLT-relations (\ref{newKLT}) that were presented
in~\cite{BjerrumBohr:2010ta} and which have the larger $(n-2)!$
permutation symmetry. It would be of interest to establish
directly, by algebraic means, the equivalence of these two
forms by use of BCJ-relations. What we have shown in this paper
is that they {\em are} identical since they both equal the
$n$-point graviton amplitude. It is a highly non-trivial task
to establish directly the relation between the $(n-2)!$
symmetric form and the $(n-3)!$ symmetric form. In principle,
we know that we need only to use BCJ-relations to establish the
connection, but the details are not simple.

One striking aspect of our proof is that it requires the
simultaneous use of a variety of different KLT-expressions,
and particular also the apparently singular version that was proven
in ref.~~\cite{BjerrumBohr:2010ta} is required.
Because these are all connected via BCJ-relations, it is
understandable why a field theory proof had to wait until these relations
had been established. Another crucial ingredient is the use of
BCFW-recursion. This shows how powerful such a recursive
technique can be, not just for explicit computation of
amplitudes, but also in a more abstract sense since they lend
themselves easily to proofs based on induction.

There are various open ends that one would like to understand
better. It would be very interesting to derive (\ref{newKLT})
from string theory. Unlike the case of~\cite{Kawai:1985xq},
this time one should fix only two closed string vertices $V_1,
V_n$. Presumably the volume of the remaining gauge symmetry
accounts, in that framework and in the field theory limit, for
the naively infinite ${1/s_{12\ldots (n-1)}}$ that needs to be
regularized by taking leg $n$ off-shell. This picture seems
intuitively reasonable, but it should be confirmed and the
details need to be figured out. Having the manifest $(n-2)!$
symmetric form, it is curious to ask if one could find manifest
$(n-1)!$ symmetric or even $n!$ symmetric forms as well.
Another formulation of KLT-relations has also been investigated
through a study of residues of poles~\cite{Other,Bern:2010ue}
and it would be interesting to consider our new more general
forms of KLT-relations in this context.

A Lagrangian prescription for KLT-factorization is still
lacking, although there has been pro\-gress in such a
direction~\cite{Bern:1999ji}. It would be interesting to
investigate this further, possibly using the methods
of~\cite{Bianchi:2008pu} now that is it clear how KLT-relations
can be completely separated from a string theory origin. There
are also alternative KLT-relations involving matter particles
that are most easily derived from the heterotic
string~\cite{Bern:1999bx} and such relations have been shown
explicitly to hold also for the corresponding effective field
theories, expanded perturbatively in $\alpha'$~\cite{EffKLT}.

Another interesting problem is the following. By direct
calculation and by using BCFW-recursion relations starting from
$M_3=A_3\W A_3$, we find $M_4=-s_{12} A(1,2,3,4)
\W A (1,2,4,3)$ after use of a BCJ-relation. As we have
stressed, the factor $s_{12}$ is important to cancel the
double pole from the product $A(1,2,3,4) \W A (1,2,4,3)$. In
fact, for the general $n$, a crucial role of the function
${\cal S}$ is to miraculously get rid of all double poles.
It is natural to ask if we can uniquely fix the function
${\cal S}$ from such a requirement.
The answer to this question may be positive. Let
us assume the result of $n$, i.e., we have
\bea M_n & = & (-)^{n+1}\sum_{\b,\a} \W A_n(n-1,
n,\a_{2,n-2},P) {\cal S}[\a|\b]_{P} A_n(P, \b_{2,n-2},
n-1,n)\,, \eea
where we have used $P$ for particle $p_1$. Going from $n$ to
$(n+1)$ we can consider $P$ as the combination of two particles
$p_1,\W p_1$. We know that for the two-particle channel, we
should have
\bea\!\!\!\!\!\! M_{n+1}& \sim &  \sum_{\b,\a} \W A_n(n-1,
n,\a_{2,n-2},\W p_1, p_1) {\cal S}[\a|\b]_{P} s_{p_1\W p_1}
A_n(p_1,\W p_1, \b_{2,n-2}, n-1,n)\nn
& \sim & \sum_{\b,\a} \W A_n(n-1, n,\a_{2,n-2},\W p_1, p_1)
{\cal S}[\a,\W p_1|\W p_1,\b]_{p_1} A_n(p_1,\W p_1, \b_{2,n-2},
n-1,n)\,.\eea
To have full $(n-2)!$ permutation symmetry, we need to sum over
all permutations between $j$ and $\W p_1$. It is then quite natural
to extend
\bea \sum_{\b,\a}{\cal S}[\a,\W p_1|\W p_1,\b]_{P} \to \sum_{\W
\b,\W \a}{\cal S}[\W\a_{p_1,2,\ldots,n-2}|\W \b_{\W
p_1,2,\dots,n-2}]\,. \eea
By this line argument, it might be that one can recursively
construct the general $n$-point KLT-relations by requiring
correct single-pole structures in addition to correct
multi-particle channels, a consistency condition that has also
been discussed in ref.~\cite{Bern:1998sv}.

There are many other directions one can follow from here. It
might be interesting to take a fresh look at string-based
techniques such as those discussed in ref.~\cite{stringbased}.
We believe that it could be useful to consider our relations
from string theory along the line of~\cite{Chen}. An obvious
question is how much can be generalized to the loop level.
Another interesting point is to see how the Grassmanian program
initiated in~\cite{Grassmanian} can be generalized to the
${\cal N}=8$ theory, perhaps by using results discussed here.
There will certainly be consistency requirements that must be
met, and it would be quite amazing to see our different
versions of KLT-relations emerge in that context.

\subsection*{Acknowledgements}
BF would like to thank for the hospitality of the Niels Bohr
International Academy and the Discovery Center where a major
part of this work was done. BF is supported by funds from
Qiu-Shi, the Fundamental Research Funds for the Central
Universities with contract number 2009QNA3015, as well as
Chinese NSF funding under contract No.10875104.

\appendix

\section{The shifting-formula for $j$ \label{appA}}
Here we provide a detailed proof of eq.~\eqref{jrel}. We
begin with the following rewriting
\begin{align}
&\sum_{\alpha} \mathcal{S}[\alpha_{i_2,i_j}|i_2,\ldots,i_j]_{p_1}
 \widetilde{A}_n(\alpha_{i_2,i_j},1,n-1,i_{j+1},\ldots,i_{n-2},n) \nonumber \\
&=\sum_{\alpha'} \bigg[ \mathcal{S}[\alpha'_{i_2,i_{j-1}},i_j|i_2,\ldots,i_j]_{p_1}
 \widetilde{A}_n(\alpha'_{i_2,i_{j-1}},i_j,1,n-1,i_{j+1},\ldots,i_{n-2},n) \nonumber \\
&+ \mathcal{S}[\alpha'_{i_2,i_{j-2}},i_j,\alpha'(i_{j-1})|i_2,\ldots,i_j]_{p_1}
 \widetilde{A}_n(\alpha'_{i_2,i_{j-2}},i_j,\alpha'(i_{j-1}),1,n-1,i_{j+1},\ldots,i_{n-2},n) \nonumber \\
&+\cdots +
\mathcal{S}[i_j,\alpha'_{i_2,i_{j-1}}|i_2,\ldots,i_j]_{p_1}
 \widetilde{A}_n(i_j,\alpha'_{i_2,i_{j-1}},1,n-1,i_{j+1},\ldots,i_{n-2},n) \bigg] \nonumber \\
&=\sum_{\alpha'} \mathcal{S}[\alpha'_{i_2,i_{j-1}}|i_2,\ldots,i_{j-1}]_{p_1} \times \bigg[
s_{1j}\widetilde{A}_n(\alpha'_{i_2,i_{j-1}},i_j,1,n-1,i_{j+1},\ldots,i_{n-2},n) \nonumber \\
&+(s_{1j}+s_{j\alpha'(j-1)})\widetilde{A}_n(\alpha'_{i_2,i_{j-2}},i_j,\alpha'(i_{j-1}),1,n-1,i_{j+1},\ldots,i_{n-2},n)
+\cdots \nonumber \\
&+ (s_{1j}+s_{j\alpha'(1)} +s_{j\alpha'(2)}+\cdots+s_{j\alpha'(j-1)})
 \widetilde{A}_n(i_j,\alpha'_{i_2,i_{j-1}},1,n-1,i_{j+1},\ldots,i_{n-2},n) \bigg]\,.
\end{align}
Using the fundamental BCJ-relation on the expression inside
$[\cdots ]$ as well as momentum conservation, we get
\begin{align}
&\sum_{\alpha'} \mathcal{S}[\alpha'_{i_2,i_{j-1}}|i_2,\ldots,i_{j-1}]_{p_1} \nonumber \\
&\times \bigg[
(s_{j(n-1)}+s_{j(j+1)}+s_{j(j+2)}+\cdots+s_{j(n-2)})
\widetilde{A}_n(\alpha'_{i_2,i_{j-1}},1,n-1,i_{j+1},\ldots,i_{n-2},i_j,n) \nonumber \\
&+(s_{j(n-1)}+s_{j(j+1)}+s_{j(j+2)}+\cdots+s_{j(n-3)})
\widetilde{A}_n(\alpha'_{i_2,i_{j-1}},1,n-1,i_{j+1},\ldots,i_{n-3},i_j,i_{n-2},n)
+\cdots \nonumber \\
&+ s_{j(n-1)}
 \widetilde{A}_n(\alpha'_{i_2,i_{j-1}},1,n-1,i_j,i_{j+1},\ldots,i_{n-2},n) \bigg] \nonumber \\
&=\sum_{\alpha'} \mathcal{S}[\alpha'_{i_2,i_{j-1}}|i_2,\ldots,i_{j-1}]_{p_1} \nonumber \\
&\times \bigg[
\frac{\widetilde{\mathcal{S}}[i_j,i_{j+1},\ldots,i_{n-2}|i_{j+1},\ldots,i_{n-2},i_j]_{p_{n-1}}}{
\widetilde{\mathcal{S}}[i_{j+1},\ldots,i_{n-2}|i_{j+1},\ldots,i_{n-2}]_{p_{n-1}}}
\widetilde{A}_n(\alpha'_{i_2,\ldots,i_{j-1}},1,n-1,i_{j+1},\ldots,i_{n-2},i_j,n) \nonumber \\
&+
\frac{\widetilde{\mathcal{S}}[i_j,i_{j+1},\ldots,i_{n-2}|i_{j+1},\ldots,i_{n-3},i_j,i_{n-2}]_{p_{n-1}}}{
\widetilde{\mathcal{S}}[i_{j+1},\ldots,i_{n-2}|i_{j+1},\ldots,i_{n-2}]_{p_{n-1}}}
\widetilde{A}_n(\alpha'_{i_2,i_{j-1}},1,n-1,i_{j+1},\ldots,i_{n-3},i_j,i_{n-2},n)
+\cdots \nonumber \\
&+\frac{\widetilde{\mathcal{S}}[i_j,i_{j+1},\ldots,i_{n-2}|i_j,i_{j+1},\ldots,i_{n-2}]_{p_{n-1}}}{
\widetilde{\mathcal{S}}[i_{j+1},\ldots,i_{n-2}|i_{j+1},\ldots,i_{n-2}]_{p_{n-1}}}
 \widetilde{A}_n(\alpha'_{i_2,i_{j-1}},1,n-1,i_j,i_{j+1},\ldots,i_{n-2},n) \bigg]\,,
\end{align}
\textit{i.e.} by multiplying with
$\widetilde{\mathcal{S}}[i_{j+1},\ldots,i_{n-2}|i_{j+1},\ldots,i_{n-2}]_{p_{n-1}}$
on both sides, we get the relation
\begin{align}
&\sum_{\alpha} \mathcal{S}[\alpha_{i_2,i_j}|i_2,\ldots,i_j]_{p_1}
\widetilde{\mathcal{S}}[i_{j+1},\ldots,i_{n-2}|i_{j+1},\ldots,i_{n-2}]_{p_{n-1}}
 \widetilde{A}_n(\alpha_{i_2,i_j},1,n-1,i_{j+1},\ldots,i_{n-2},n) \nonumber \\
&= \sum_{\alpha'} \mathcal{S}[\alpha'_{i_2,i_{j-1}}|i_2,\ldots,i_{j-1}]_{p_1} \nonumber \\
&\times \bigg[
\widetilde{\mathcal{S}}[i_j,i_{j+1},\ldots,i_{n-2}|i_{j+1},\ldots,i_{n-2},i_j]_{p_{n-1}}
\widetilde{A}_n(\alpha'_{i_2,i_{j-1}},1,n-1,i_{j+1},\ldots,i_{n-2},i_j,n) \nonumber \\
&+
\widetilde{\mathcal{S}}[i_j,i_{j+1},\ldots,i_{n-2}|i_{j+1},\ldots,i_{n-3},i_j,i_{n-2}]_{p_{n-1}}
\widetilde{A}_n(\alpha'_{i_2,i_{j-1}},1,n-1,i_{j+1},\ldots,i_{n-3},i_j,i_{n-2},n)
+\cdots \nonumber \\
&+\widetilde{\mathcal{S}}[i_j,i_{j+1},\ldots,i_{n-2}|i_j,i_{j+1},\ldots,i_{n-2}]_{p_{n-1}}
 \widetilde{A}_n(\alpha'_{i_2,i_{j-1}},1,n-1,i_j,i_{j+1},\ldots,i_{n-2},n) \bigg]\,.
\end{align}
We can now add all the permutations of leg $j+1,\ldots,n-2$. On
the left-hand side we get
\begin{align}
&\sum_{\alpha,\beta} \mathcal{S}[\alpha_{i_2,i_j}|i_2,\ldots,i_j]_{p_1}
\widetilde{\mathcal{S}}[i_{j+1},\ldots,i_{n-2}|\beta_{i_{j+1},i_{n-2}}]_{p_{n-1}} \nonumber \\
&\times \widetilde{A}_n(\alpha_{i_2,i_j},1,n-1,\beta_{i_{j+1},i_{n-2}},n)\,,
\end{align}
and on the right-hand side, since we have all permutations of
$i_j$ with the $\{j+1,\ldots,n-2\}$ set, we can write it as all
permutations of $\{i_j,i_{j+1},\ldots,i_{n-2}\}$
\begin{align}
&\sum_{\alpha',\beta'} \mathcal{S}[\alpha'_{i_2,i_{j-1}}|i_2,\ldots,i_{j-1}]_{p_1}
\widetilde{\mathcal{S}}[i_j,i_{j+1},\ldots,i_{n-2}|\beta'_{i_j,i_{n-2}}]_{p_{n-1}} \nonumber \\
&\times \widetilde{A}_n(\alpha'_{i_2,i_{j-1}},1,n-1,\beta'_{i_j,i_{n-2}},n)\,,
\end{align}
and we finally get the relation
\begin{align}
&\sum_{\alpha,\beta} \mathcal{S}[\alpha_{i_2,i_j}|i_2,\ldots,i_j]_{p_1}
\widetilde{\mathcal{S}}[i_{j+1},\ldots,i_{n-2}|\beta_{i_{j+1},i_{n-2}}]_{p_{n-1}} \nonumber \\
&\times \widetilde{A}_n(\alpha_{i_2,i_j},1,n-1,\beta_{i_{j+1},i_{n-2}},n) \nonumber \\
&=\sum_{\alpha',\beta'} \mathcal{S}[\alpha'_{i_2,i_{j-1}}|i_2,\ldots,i_{j-1}]_{p_1}
\widetilde{\mathcal{S}}[i_j,i_{j+1},\ldots,i_{n-2}|\beta'_{i_j,i_{n-2}}]_{p_{n-1}} \nonumber \\
&\times \widetilde{A}_n(\alpha'_{i_2,i_{j-1}},1,n-1,\beta'_{i_j,i_{n-2}},n)\,.
\end{align}
This ends our proof of eq.~\eqref{jrel}. This then also proves
the $j$-independence of eq.~\eqref{KLT-bern}.



\end{document}